\documentclass[a4paper,11pt]{article}
\pdfoutput=1 

\usepackage{jinstpub} 

\title{\boldmath Fabrication and beam test of a silicon-tungsten 
  electromagnetic calorimeter}

\author[a,1]{Sanjib Muhuri,\note{Corresponding author.}}
\author[b]{Sourav Mukhopadhyay,} 
\author[b]{Vinay B. Chandratre,}
\author[a,e]{Tapan K. Nayak,}
\author[a]{Sumit Kumar Saha,} 
\author[a]{Sanchari Thakur,} 
\author[a]{Rama N. Singaraju,}
\author[a]{Jogender Saini,}  
\author[c]{Anthony van den Brink,}
\author[d]{Tatsuya Chujo,}
\author[a]{Rajendra Nath Patra,}
\author[c,e]{Marco van Leeuwen,}
\author[a]{Shuaib Ahmad Khan,}
\author[b]{Menka Sukhwani,}
\author[c]{Gert-Jan Nooren,}
\author[c]{Thomas Peitzmann}

\affiliation[a]{Variable Energy Cyclotron Centre, HBNI, Kolkata - 700064, India}
\affiliation[b]{Bhabha Atomic Research Centre, Electronics Division, HBNI,
  Trombay, Mumbai - 400085, India}
\affiliation[c]{Institute for Subatomic Physics, Utrecht University and Nikhef, PO Box 80000, 3508TA Utrecht, The Netherlands}
\affiliation[d]{University of Tsukuba, Tsukuba, Ibaraki 305-8577, Japan}
\affiliation[e]{CERN, Geneva 23, Switzerland.}
\emailAdd{sanjibmuhuri@vecc.gov.in}

\abstract{A silicon-tungsten (Si-W) sampling calorimeter, consisting of 19
alternate layers of silicon pad detectors (individual pad area of
1~cm$^2$) and tungsten absorbers (each of one radiation length), has
been constructed for measurement of electromagnetic showers over a
large energy range. The signal from each of the silicon pads is
readout using an 
ASIC  with a dynamic range from $-300$~fC to $+500$~fC. Another ASIC
with a larger dynamic range, $\pm 600$~fC has been used as a test study.  
The calorimeter was exposed to pion and electron beams at the CERN Super Proton Synchrotron (SPS)
to characterise the response to minimum ionising particles
(MIP) and showers from electromagnetic (EM) interactions. Pion beams
of 120 GeV provided baseline measurements towards the understanding of
the MIP behaviour in the silicon pad layers, while electron beams of energy from 
5 GeV to 60 GeV rendered detailed shower profiles within the calorimeter. 
The energy deposition in each layer, the longitudinal shower
profile, and the total energy deposition have been measured for each incident
electron energy. Linear behaviour of the total measured energy ($E$) with that
of the incident particle energy ($E_{0}$) ensured satisfactory calorimetric
performance. For a subset of the data sample, selected based on the
cluster position of the electromagnetic shower of the 
incident electron, the dependence of the measured energy resolution on 
$E_{0}$ has been found to be $\sigma/E = (15.36/\sqrt{E_0(\mathrm{GeV)}} \oplus 2.0) \%$.}

\keywords{Calorimeter, electromagnetic shower, silicon pad detectors, longitudinal profile, energy resolution.}

\begin{document}
\maketitle
\flushbottom

\section{Introduction}
\label{sec:intro}

Electromagnetic sampling calorimeters, consisting of alternating
layers of thin tungsten absorber and granular silicon sensors have
been under active study for the last two decades for applications in
high-energy physics experiments. In the present era of large-scale
experiments at the Large Hadron Collider (LHC), Relativistic Heavy Ion
Collider (RHIC), Japan Proton Accelerator Research Complex (J-PARC),
Facility for Antiproton and Ion Research (FAIR) and proposed
International Linear Collider (ILC), the field of calorimetry is
constantly evolving, adapting to new technologies and new ideas. 
An example of a physics study using such a calorimeter is to constrain the low-x parton structure
of protons and nuclei via the measurement of direct photons at
the LHC. This requires, in particular, the discrimination of single
photons from the dominating number of decay photons arising from
neutral pions ($\pi^0$) and $\eta$. Measurement of the decay photons
and reconstruction of the parent particles need a
calorimeter with fine granularity with excellent two-shower separation
capability~\cite{TP,MK,Calorimeter-1,TB-paper}. Thus one of the primary objectives of the
present calorimeter is to provide position and energy
measurements of photons and thereby $\gamma$-$\pi^0$ discrimination within a
large momentum range ($\approx$ up to 50~GeV/c). Segmented sampling calorimeters with alternate
passive and active media 
have the advantage to provide more differential measurements of particle showers
~\cite{sicapo-1, sicapo-2,sicapo-3,wizard,opal,calice-1,
calice-2,calice-3,calice-4,phenix,phenix1,strom,cluster,Calorimeter-2,tapan-x}
with with an energy resolution of the order of a few \% at high energy. Segmentation
in both longitudinal and transverse directions helps in particle
identification, providing better $\gamma$-$\pi^0$ and $\gamma$-charged hadron
discrimination compared to other available calorimetric options. A highly segmented calorimeter can also offer advantages for jet measurements when using particle flow algorithms~\cite{PFalgo}.

Silicon sensors are most suitable as detection medium in high-energy
physics experiments because of their
good charge collection
efficiency, fast response time and low bias voltage to achieve full
depletion. Low
sensitivity to magnetic fields makes silicon attractive for
experiments with large ambient magnetic fields. As a converter
material, tungsten is most suitable because of the small Moliere radius, which facilitates the containment of the energy of the incident particles fully in relatively
small volumes. Sampling calorimeters using silicon and tungsten
have been proposed for upgrades of the ALICE and CMS experiments at the LHC as well as in
other contemporary experiments. Based on GEANT simulation
studies~\cite{cluster} and a feasibility test with only a few
layers~\cite{TB-paper},  a 19 layer prototype Si-W calorimeter has
been fabricated and tested with pion and electron beams at the
CERN-SPS. The tungsten absorbers are of one radiation length ($X_0$)
thickness, which results in a total depth of the calorimeter of about
19$X_0$. 

In this article, we present the design of the calorimeter
along with readout electronics and discuss the test beam
results. In section~2, we discuss the design and optimisation of the
sampling calorimeter. In section~3, we present  the 
calorimeter setup and electronics readout. The test beam setup at SPS
with trigger logic as well as the electronics noise for each
detector pad are discussed in section~4. 
Results for pions incident on
the calorimeter are presented in section~5. The response of the
silicon layers to electron beams at different energies is presented
in section~6 and compared to simulated results. In section~7, we
present the results for energy
resolution of the calorimeter. The article
concludes with a summary in section~8.

\section{Design and optimisation of the Si-W Calorimeter}

\begin{figure}[!th]
\centering \includegraphics[width=0.7\textwidth]{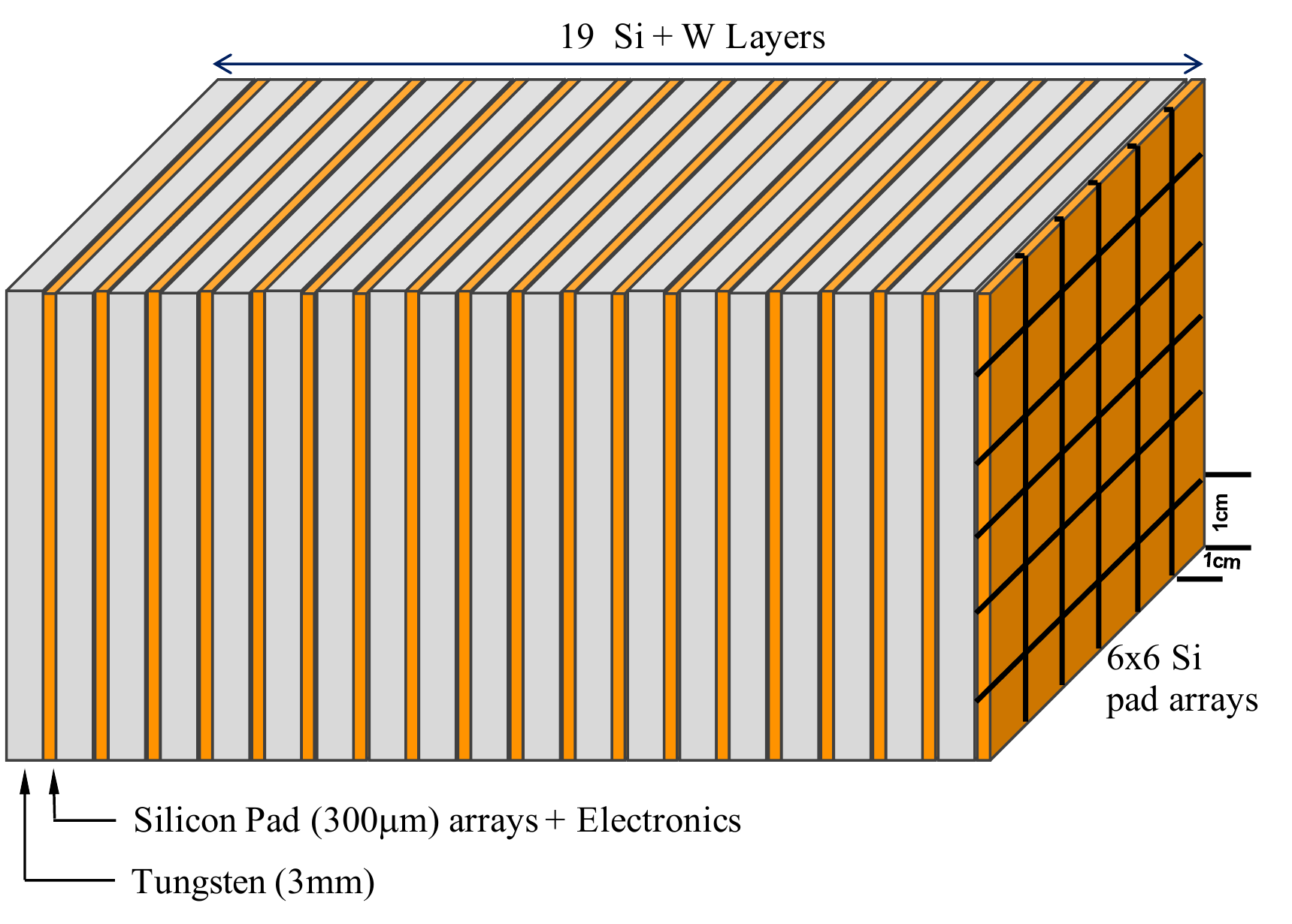}
\caption{The components of a silicon-tungsten sampling calorimeter consisting of 
19 alternating layers of silicon pad detectors (individual pad area of 1~cm$^2$)
and tungsten plates (1~X$_0$ thick).}
\label{fig:calorimeter}
\end{figure}

Geometry optimisation and characterisation of a sampling Si-W
electromagnetic calorimeter have been carried out using the
GEANT4~\cite{geant4_1,geant4_2} geometry package. The main goal is to
optimise the geometry to achieve a good position and energy resolution
in a high particle density environment. In total 19 layers, each
consisting of silicon pad arrays and tungsten absorber have been
implemented. The silicon pads are of 1~cm$^2$ area and 300~$\mu$m in
thickness. A gap of 3mm was kept for electronics and readout behind
each silicon pad layer. High purity tungsten plates of density
19.3gm/~cm$^3$ and dimension 10~cm$\times$10~cm with 0.35~cm thickness
have been placed alternately with silicon layers.  The calorimeter setup is shown in
figure~\ref{fig:calorimeter}. 

In high-energy collisions, measurements of direct photons (which are
not coming from hadronic decays) are of particular interest since they
directly probe partonic processes. Direct photons originate directly
from hard interactions, such as quark-anti-quark annihilation,
quark-gluon Compton scattering, and higher order processes. In
experiments, the large background of photons from hadronic decays is
partially removed by selections and the remaining background is
subtracted to determine the direct photon cross section. One of the
selection cuts that is frequently employed is a so-called isolation
cut that selects photons that are not accompanied by additional
momentum flow; such a cut rejects both decay photons from quark and
gluon jets as well as so-called fragmentation photons, which are
produced as radiation off a high-energy quark. 
The rejection and subtraction of the contributions from hadronic
decays requires an accurate reconstruction of
$\pi^0$  and $\eta$ from its decay photons.
The energy deposition is measured in each individual pad sensor 
of each silicon layer. The photon showers are reconstructed in the
silicon layers, which provide the position of the shower center, shower shape, and total energy deposition.

With the increase of the parent $\pi^0$ energy, the opening angle between the
decay photons decreases, which is not possible to measure with the
above configuration of the calorimeter. To improve the measurement of two-shower
separations, a modified setup has been planned in future by
including two or three very high granularity silicon pixel (of size 1~${\rm mm}^2$) layers at depths around the shower maximum. 
Studies with a high-granularity calorimeter using pixel sensors are reported elsewhere~\cite{maps}.

\section{Calorimeter setup}

\begin{figure}[!th]
\centering \includegraphics[width=0.7\textwidth]{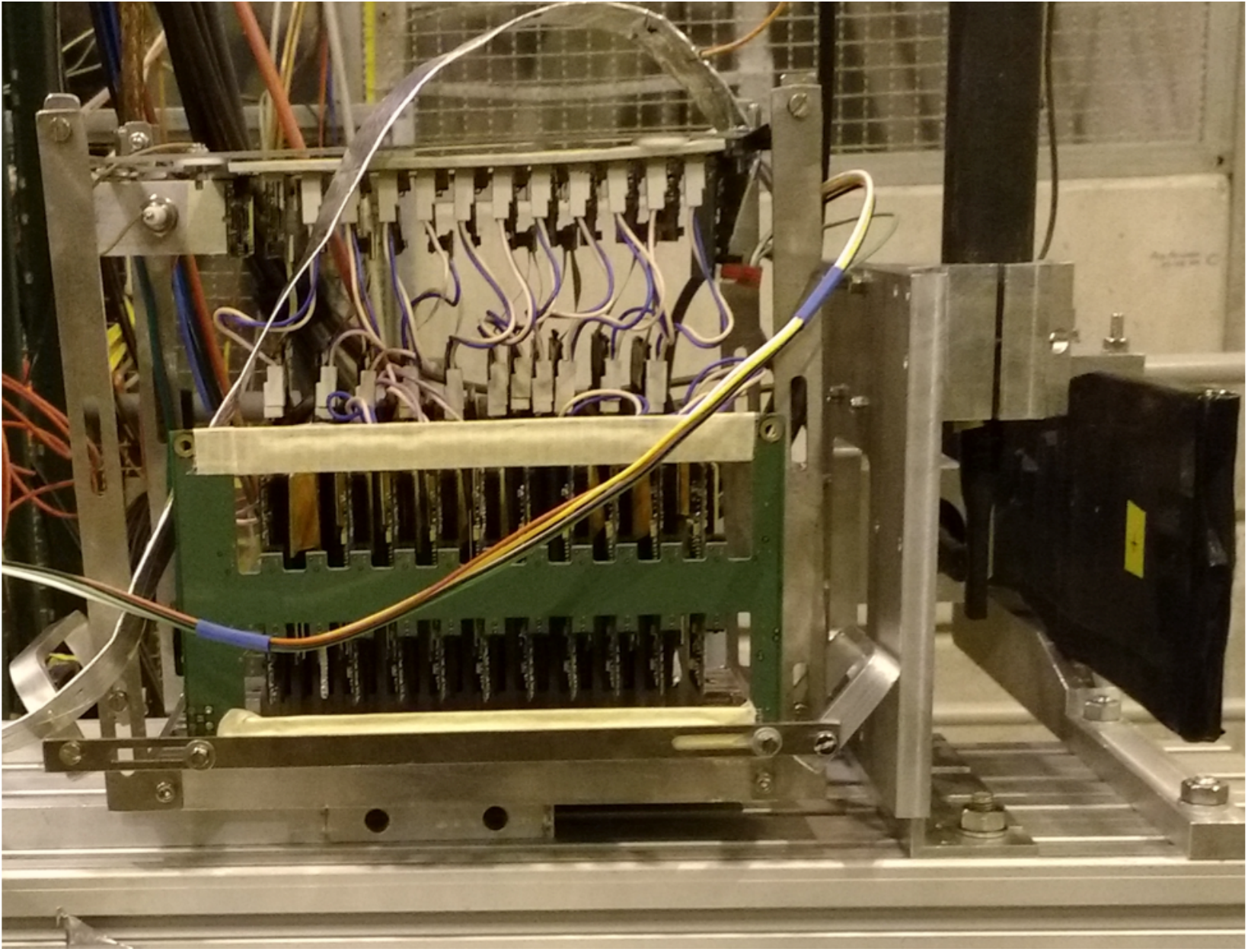}
\caption{Experimental arrangement of the Si-W prototype calorimeter with 19
  layers of silicon pad detectors arrays and tungsten plates along
  with associated electronics for the beam test at CERN-SPS.}
\label{setup}
\end{figure}

Figure ~\ref{setup} shows a photograph of the calorimeter setup. A
custom-designed mechanical arrangement for holding the tungsten plates
and silicon pad detectors properly in place has been constructed. The
average gap between two consecutive tungsten layers was kept at 0.4~cm
to accommodate a silicon detector mounted on a 0.08~cm thick PCB and
the associated electronics.

\subsection{Silicon pad array}

\begin{figure}[!th]
\centering \includegraphics[width=0.7\textwidth]{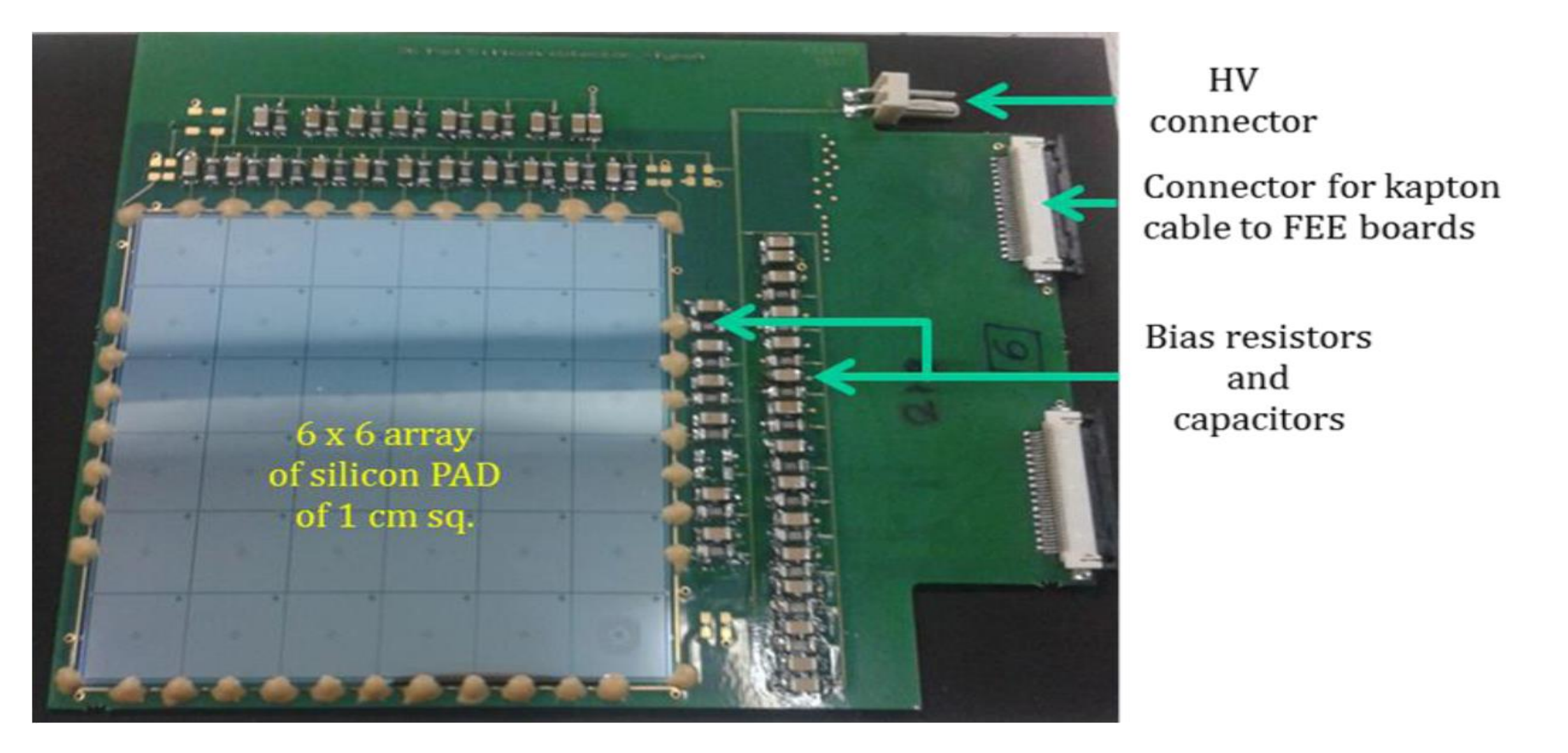}
\caption{Silicon detector array consisting of $6\times  
  6$ pads (each of 1~cm$^{2}$ area). }
\label{silicon_pad}
\end{figure}

6$\times$6 arrays of silicon pad detectors with 1~cm$^{2}$ individual
pad size are developed on 4-inch wafers with 36 pads per detector
layer. The detectors were fabricated~\cite{bel} on $300\mu$m-thin 
n-type FZ (Float-Zone) wafers with $<111>$ crystal orientation. 
Figure~\ref{silicon_pad} shows the photograph of such a silicon detector layer. 
The starting resistivity of the wafer is 4-5 k$\Omega$-cm. The p-n junctions are 
created by implanting p-type impurity from the top side. 
High breakdown voltage and low leakage current are the essential
features of silicon pad detectors for HEP experiments to compensate
the performance degradation due to long term operation under large
particle fluence. 
The design of the
silicon pad is carried out with multiple FGR (Floating Guard Rings)
and MO (Metal Overhang) over the individual pad to reduce the
electric field crowding at the junction edge near the surface and thus
improving the breakdown voltage ~\cite{pad1}. Moreover gettering
techniques and double implantation steps for the backside ohmic
contact have been incorporated to reduce the leakage current of the
large area silicon detectors. Alignment sites are placed in each pad
for checking the integrity of an individual pad connection. The
silicon pad detectors are designed with a conventional wire bonding
packaging option. Due to limitations of the foundry, all the
interconnections have to make use of a single metal layer. The
distances and routing have been optimized to ensure minimum cross-talk
among the pads and a minimal dead area.

The silicon pads are die-attached and mounted on a
0.8~mm-thick four-layer PCB with a silver conductive epoxy of
resistivity 0.006 $\Omega$-cm followed by gold wire bonding from each
pad to the PCB. The fabricated and packaged detectors have less than
10nA/~cm$^{2}$ leakage current with a breakdown voltage of more than
500~V. With full depletion achieved around 45~V, each pad has 45 -
50~pF/~cm$^{2}$ detector capacitance.

\subsection{Readout electronics}

The detector signals were readout using front-end electronics boards
making use of two different readout ASICs, called MANAS and
ANUSANSKAR. Both ASICs have 16 low-noise pulse processing
channels. Each channel is comprised of Charge Sensitive Amplifier,
semi Gaussian shaper, track and hold and analog serial data
readout. The linear dynamic range of MANAS is from $+500$~fC to
$-300$~fC, while ANUSANSKAR ASIC has a dynamic range of $\pm 600$~fC. The
results presented in this report  are based on data taken with MANAS
as readout ASIC.

In each pad layer, 32 pads out of 36 are connected to the
readout electronics, excluding the four corner pads. 
This is because the readout board contains 64 channels and so for ease of
connection two pad layers are connected to one readout board. In total 
19$\times$32 electronics channels are connected to the pad detectors. 
The beam was incident at the center of one of the central pads. 
As the transverse shower spread is expected to be about 1~cm, the exclusion of the corner pads
is not expected to significantly affect the measurement of the shower properties.


\section{Test beam setup at CERN-SPS}

The Si-W calorimeter has been tested using the H6 beamline 
facility~\cite{H6-Beam} of CERN SPS. In this beamline, secondary 
particle beams provide hadrons, electrons or muons at momenta between 
5~GeV/$c$ and 205~GeV/$c$, depending on the setup. Three distinct 
modes of operation are possible in this beamline, which are the 
high-resolution mode, the high transmission mode and the filter mode 
or test-beam mode. Here, the filter mode has been used for the beam 
operation, which gives a large range of incident energies.  Maximum 
intensities achievable with the H6 beamline for $10^{12}$ primary 
protons at 400 GeV/c are $10^{8}$ and $4\times10^{7}$ for $\pi^{+}$
and $\pi^{-}$, respectively at 150 GeV/c. For optimising the purity of 
the electron beam, a Cherenkov detector is used as a part of the beamline setup.

\begin{figure}[!th]
\centering 
\centering \includegraphics[width=0.99\textwidth]{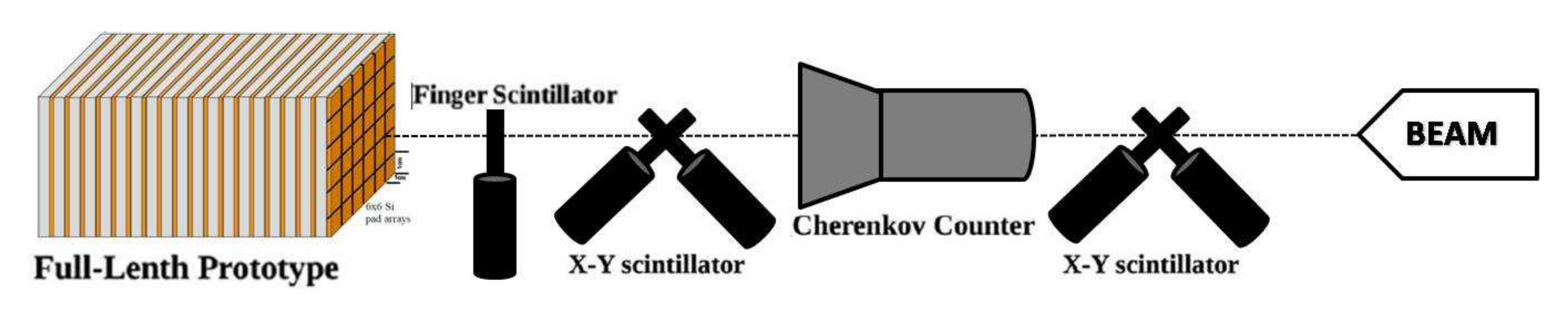}
\caption{Schematic setup of the Si-W calorimeter along with the trigger arrangement for beam tests at CERN-SPS. }
\label{schema}
\end{figure}
 
The calorimeter was placed in the H6 beam hall on an adjustable table,
capable of moving both in horizontal and vertical directions. The 
movements could be controlled remotely from the counting room. All the 
detector layers were properly shielded against both electromagnetic 
interference and ambient light to improve the signal to noise 
ratio. The readout ASICs placed close to silicon detectors, were 
coupled to the CROCUS data acquisition via MARC ASIC 
interface~\cite{muon,pmd}. The data were sent to the counting room 
through optical fiber cables.

A dedicated trigger setup, with two 
pairs of scintillator paddles and a finger scintillator, were used to 
define the beam. The trigger arrangement is shown in 
figure~\ref{schema}. The two pairs of scintillator paddles define the 
$X-Y$ position of the incoming beam within 1~cm$^2$ area whereas the 
small finger scintillator (3mm$\times$3mm) helps to position the beam 
mostly within a single silicon pad. The coincidence signal from the 
scintillator paddles and the finger scintillator was used to select 
the beam. Electrons were filtered using the Cherenkov detector in 
addition to the scintillators. 

\begin{figure}[!th]
\centering \includegraphics[width=0.7\textwidth]{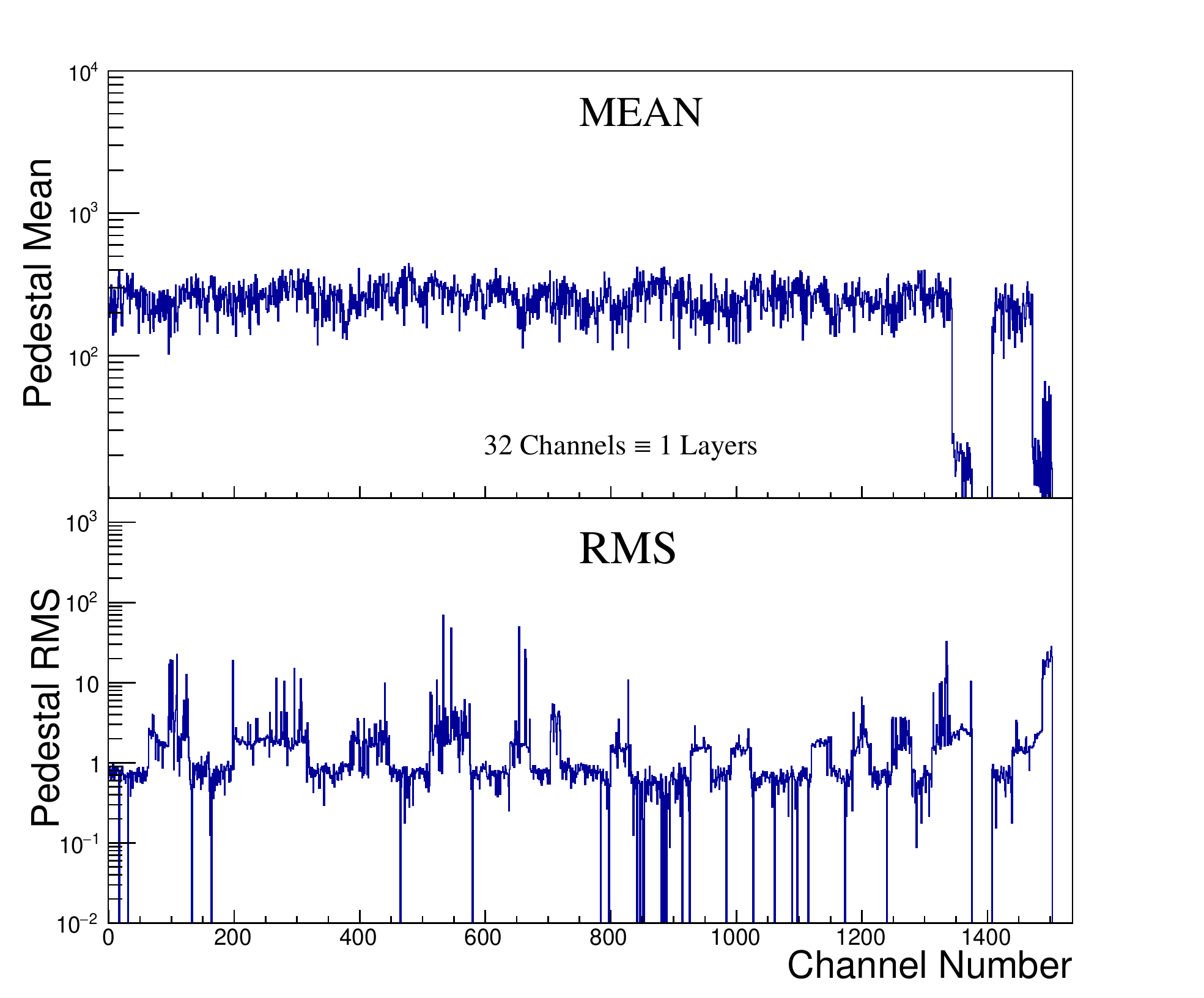}
\caption{Pedestal mean (upper panel) and RMS (lower panel) 
as a function of detector channel number. 
Groups of 32~channels correspond to individual layers along the depth 
of the calorimeter. 
}
\label{pedrms}
\end{figure}

The electronics noise arising both from the detector and the
readout has been measured for each channel of the calorimeter without the beam, which
gives the pedestal of the ADC distribution.  For a given channel, the
pedestal is defined by the
mean value of the ADC and the corresponding RMS deviation.
In Fig.~\ref{pedrms}, we plot the mean
and RMS of the pedestal for each silicon pad (each electronics
channel). 
The mean values were consistently in the vicinity of 250 ADC counts and were
stable with time. The RMS (representing the noise) were between 1.5 to
2.5 ADC counts.  Some fluctuations have been observed in the RMS
distributions in some of the 
channels, where RMS values are either very large (noisy channels) or
have values close to zero (dead channels). This can arise because of
processing errors at the foundry level or because of faulty
electronics. During the next iteration of the detector, care will be taken to
minimize these faults. There were 65 such
noisy/dead cells out of the 608 total. 
For calculating the total response from the EM showers, these channels are assigned ADC values
which are averages of their adjacent channels.

In the present analysis, each pad response is obtained by subtracting the pedestal (mean) from the measured ADC counts. 
The pad response corresponds to the amount of deposited energy in the
silicon pad.
A sum of the pad response over the entire layer gives the response of
the silicon pad layer. 

\section{Response of the calorimeter to pion beam }

\begin{figure}[!th]
\centering \includegraphics[width=0.55\textwidth]{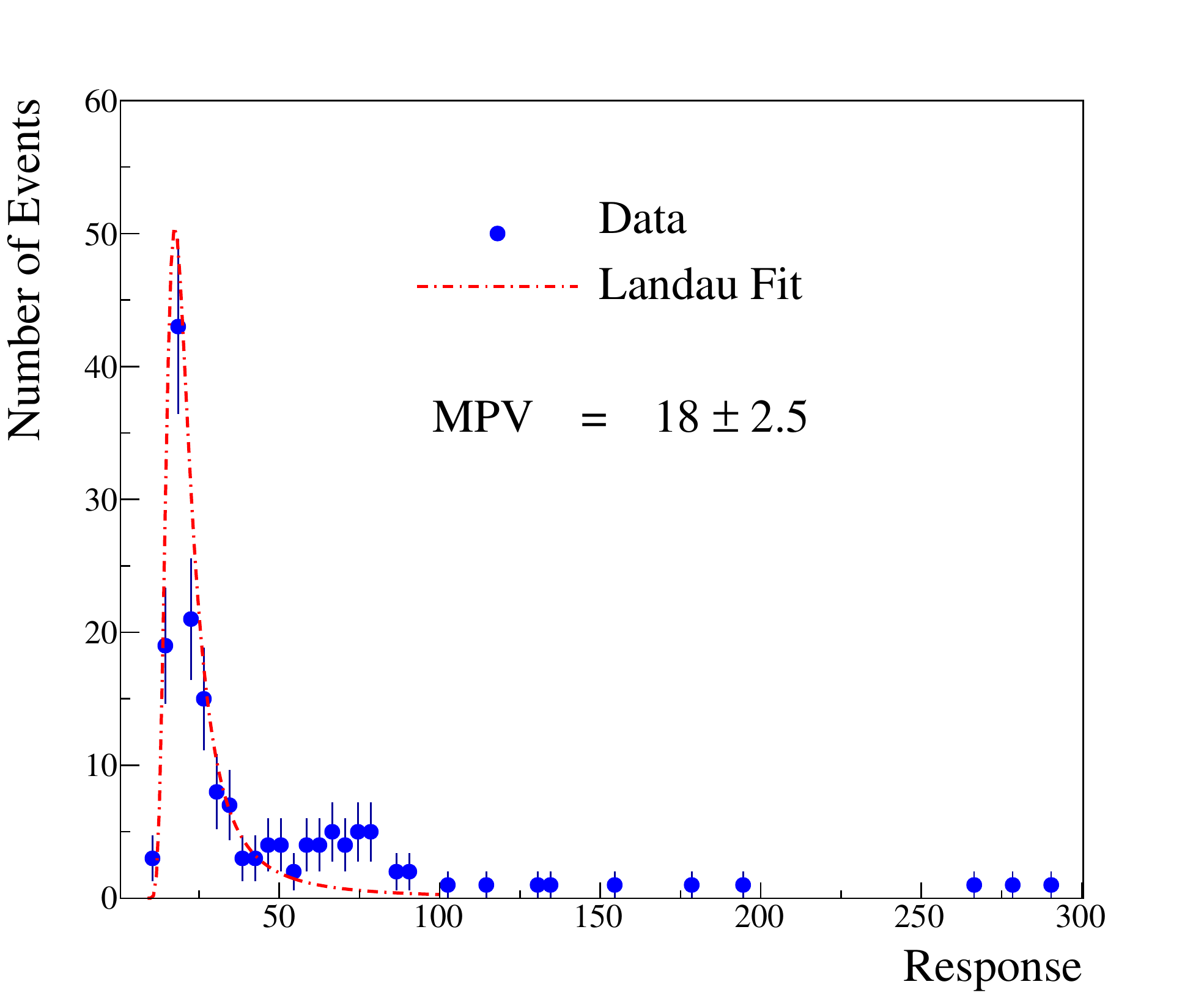}
\centering \includegraphics[width=0.55\textwidth]{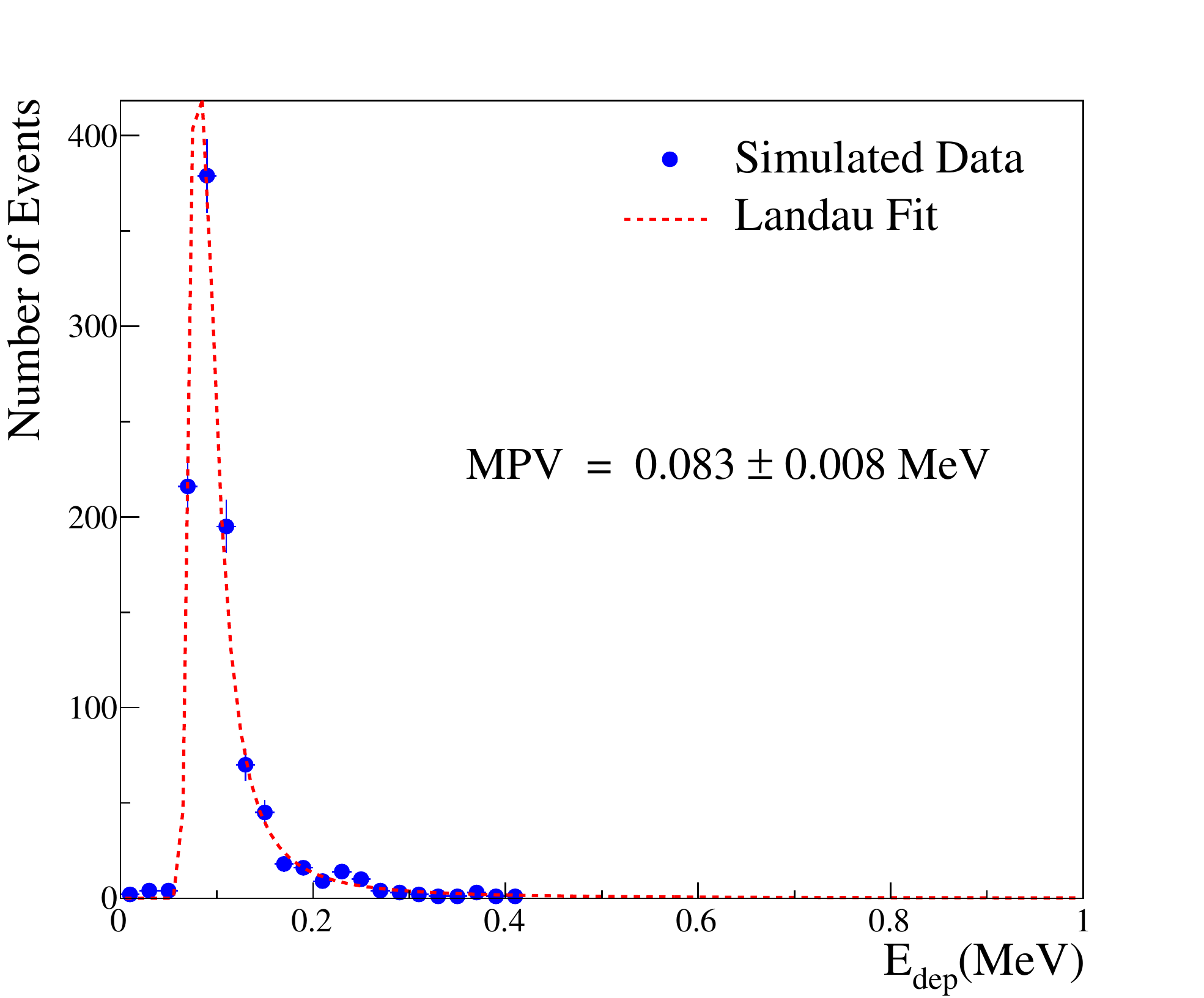}
\caption{The top panel shows the measured response of silicon pad detectors to 
120~GeV pions. The statistical uncertainties are small and within the marker size. 
The bottom panel gives the energy deposition in MeV from the simulated data. 
Both the distributions are fitted with a Landau function. 
}
\label{pion}
\end{figure}

The Si-W calorimeter has been exposed to pion beams at an incident
energy of 120GeV. Because the finger scintillator is placed at the
center of one of the silicon pads, most of the time only this center
pad is hit. The distribution of single layer response as measured by
the second silicon detector layer is shown in the upper panel of
figure~\ref{pion}. The data points are plotted after pedestal
subtraction. Minimum ionising particles are selected by selection events where the total observed signal in the detector is below 2000 ADC counts, which is about 5 times the most probably response for minimum ionising particles. The histogram is fitted with a Landau distribution, which gives a most
probable value (MPV) of 18.0$\pm$2.5. This is similar to what was
obtained in an earlier test beam~\cite{TB-paper} without any material
in front of the silicon detector. The corresponding value of deposited
energy from the simulation is 0.083~MeV as shown in the lower panel of
figure~\ref{pion}.

\section{Response of the calorimeter to electrons}  

Electrons incident on the Si-W calorimeter produce electromagnetic
showers which leave hits distributed among pads in each of the
layers. The response in each layer has been studied to understand the
shower profile along the longitudinal direction.
In general, showers induced by incident electrons produce signal in
several pads; the number of pads depends on the incident
energy. 

\subsection{Energy deposition in different layers}

\begin{figure}[!th]
\centering \includegraphics[width=0.7\textwidth]{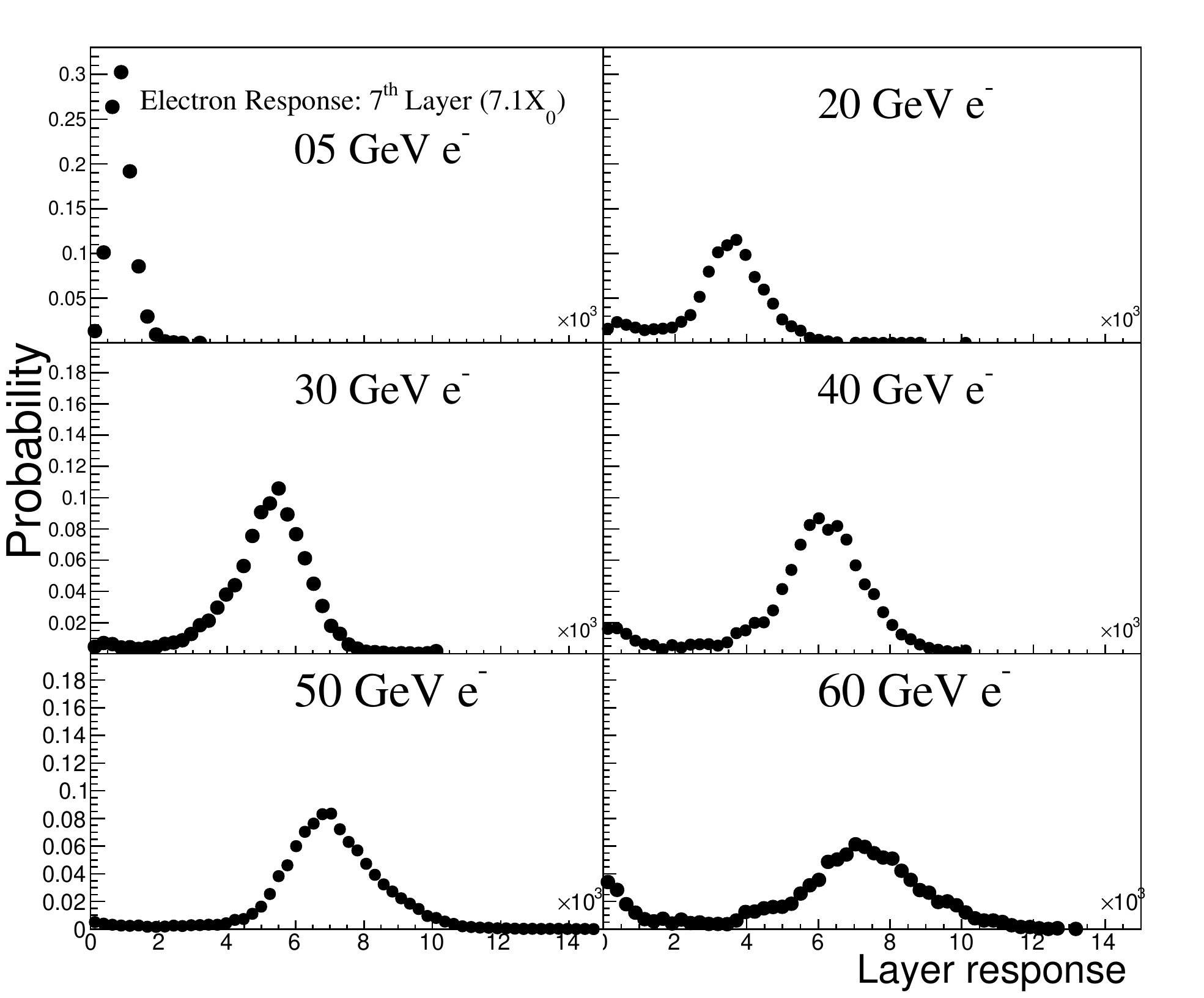}
\caption{
Measured response on the 7$^{th}$~layer 
of the Si-W calorimeter for 5, 20, 30, 40, 50 and 60~GeV electrons. 
The statistical uncertainties are small, and within the size of the marker. 
 Gaussian fits to the curves give the mean response. 
}
\label{electron}
\end{figure}

The distribution of the layer response for the
7$^{th}$ layer are presented in figure~\ref{electron} for incident
electrons of six different energies, $E_0 =$ 5, 20, 30, 40, 50 and
60~GeV. The ordinate is expressed in terms of probability, where the
counts are normalised to the number of events. These measured response
spectra have approximate  Gaussian shapes, so mean values are
extracted using Gaussian fits. The results of these fits are analysed
below to obtain the longitudinal shower profile. Similar fits are
performed to the sum of the signals to obtain the total response and
the energy resolution of the prototype calorimeter.

\subsection{Longitudinal shower profile}

\begin{figure}[!th]
\centering 
\centering \includegraphics[width=0.55\textwidth]{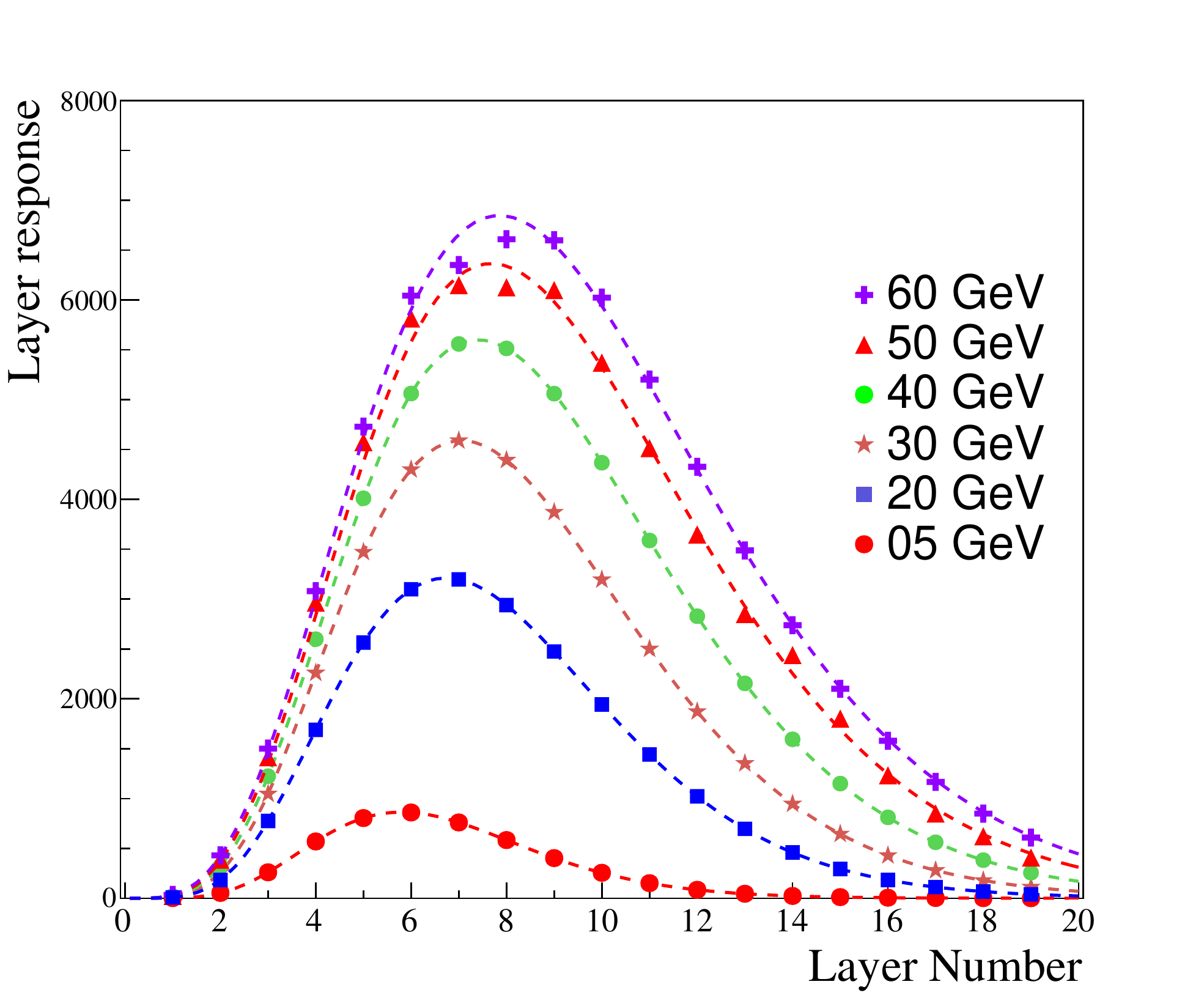}
\caption{(Colour Online) Longitudinal shower profiles of the Si-W calorimeter for 5, 20, 30,
40, 50 and 60~GeV electrons as a function of the silicon layer along
the depth of the calorimeter. The statistical uncertainties are small
and within the size of the marker. The data 
  are fitted with an empirical curve as given in equation~\ref{eq1}.} 
\label{long_data}
\end{figure}

The longitudinal profile of the EM shower is obtained by plotting the
measured mean of the response for each layer along the depth of the
calorimeter, i.e. as a function of the layer
number. Figure~\ref{long_data} shows the longitudinal shower profiles
for different incident electron energies. Here each layer corresponds
to 1.013$X_0$. For all incident energies, the response increases as a
function of layer number until it reaches a maximum value and then
decreases gradually. The shower profile can be described by the
empirical formula~\cite{longprof}, 
\begin{equation}
\label{eq1}
\frac{dR}{dt} = R_{0} \cdot t^{a} \cdot e^{-bt},
\end{equation}
where $R_{0}$ is a normalisation parameter, $t = z/X_0$ is the
thickness of the absorber in front of the layer in units of radiation
length, and $a$ and $b$ are fit parameters describing the shape. The
longitudinal shower profile has been fitted using this formula and the
fitted curves are superimposed on the measured data points. These
curves describe the data well. At higher energies, flatter regions are
seen around the shower maximum position, which may be caused by
saturation of the ADC of the MANAS readout.

\begin{figure}[!th]
\centering \includegraphics[width=0.55\textwidth]{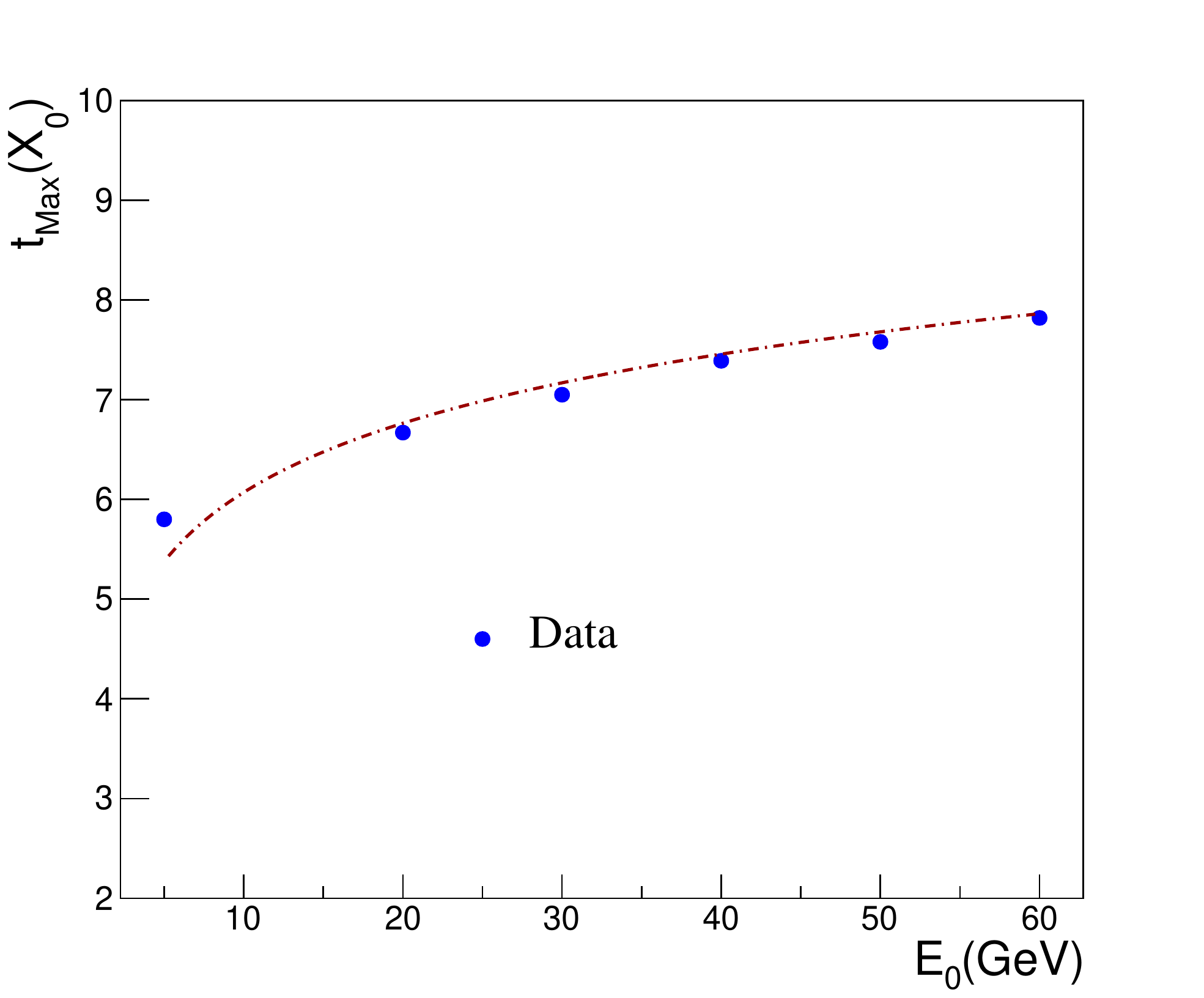}
\caption{Position of the shower maximum in the Si-W
  calorimeter as a function of incident electron energy. The
  statistical errors are small and within the size of the marker.
The line corresponds to the function given in equation \ref{eq2}. }
\label{electron2}
\end{figure}

The position of the shower maximum depends on the incident electron energy and can be expressed in terms of the fit parameters, $a$ and $b$, as $t_{\mathrm{max}} = b/a$. The extracted values of the shower maximum have been plotted as a function of the incident energy in figure~\ref{electron2}. The depth of the shower maximum position increases with the increase of the incident energy. The shower maximum position can also be described by the empirical formula~\cite{longprof}: 
\begin{equation}
\label{eq2}
 t_{\mathrm{max}} =\ln\left(\frac{E_{0}}{E_{C}}\right) + C_{e,\gamma},
\end{equation}
where $E_{C}$ is the critical energy, at which the electron ionisation losses and bremsstrahlung losses become equal, given by $E_C = 550 \mathrm{MeV}/Z_{\mathrm{eff}}$. 
$C_{e,\gamma}$ is a parameter, which takes different values for electrons and photons, here we use $C_e = -1.0$. The empirical formula is shown as the dashed curve in figure~\ref{electron2} and is seen to describe the data reasonably well.

\section{Total energy response of the calorimeter}

The total energy deposition in the calorimeter can be obtained by summing the 
pedestal-subtracted energy response (ADC) of the individual pads of
the silicon layers. The details of the electromagnetic shower
propagation inside the calorimeter can be better understood in terms
of the parameters associated with the calorimetric responses to the energy
deposition such as calibration, and energy resolution. However, it is to be noted that non-uniformity in the detector 
and the gaps in between the silicon wafer and
tungsten do influence the response when showers traverse
these regions~\cite{calice-1}. In the absence of a direct estimation of
such effects, we have studied the cluster position of the electromagnetic shower by projecting all the
pad level signals onto a single layer of the calorimeter and selecting
events which satisfy a criterion set on
the position of the cluster centres on an event-by-event basis. The
cluster centres ($X_{\mathrm{C}},Y_{\mathrm{C}}$) are determined by calculating the centre of gravity with logarithmic weights, to improve the position resolution.
The signals are first summed over all layers for each transeverse pad
position and. The centre of gravity is then calculated by summing over
the pads:

\begin{equation}
\label{clspos}
X_{\mathrm{C}} =\frac{\sum_{\rm i} w_{\rm i}x_{\rm
     i}}{\sum_{\rm i} w_{\rm i}},
\end{equation}

where $i$ is the pad index, $x_{i}$ is the transverse position (in $x$) and $w_{i} = \mathrm{max}\left(0,\left[w_{0} + \ln\left(\frac{E_{i}}{E_{\rm T}}\right)\right]\right)$ is the weight, which depends on 
 the energy deposition $E_{i}$ in the pad and the total shower energy $E_{\rm T}$. The $y$-coordinate of the centroid $Y_{\mathrm{C}}$ is calculated in the same way, using $y_i$ instead of $x_i$.
Events are selected to have their shower position in the active area that is selected by the trigger scintillator within a region of uniform response; the range $1.8<X_{\mathrm C}<2.2 \, \mathrm{cm}$ and
$2.5<Y_{\mathrm C}<2.8 \, \mathrm{cm}$ is selected. The origin of the 
coordinate system is the lower left corner pad of the
layer. It was verified that this selection does not bias the energy response significantly.

\subsection{Response to different incident energies}

\begin{figure}[!th]
\centering \includegraphics[width=0.55\textwidth]{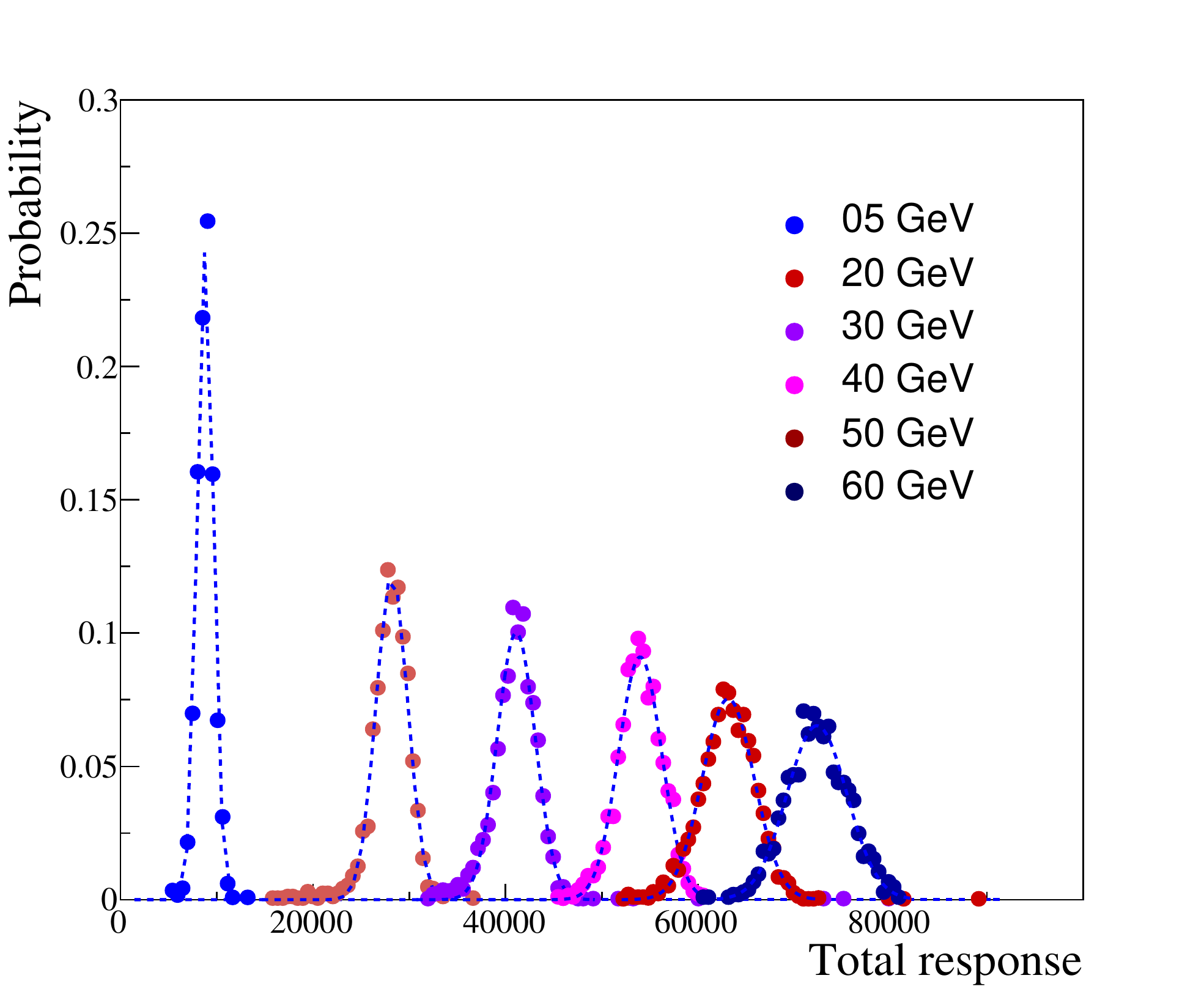}
\caption{(Colour Online) Response of the calorimeter for electrons,
  calculated using the selected events depending on the cluster centre position. The error bars represent statistical uncertainties. 
}
\label{AllEnerElecSubsampleMethod}
\end{figure}

Figure~~\ref{AllEnerElecSubsampleMethod} 
shows distributions of the total energy deposition in the calorimeter
for six incident energies. Gaussian shapes have been observed for each
incident energy and the
distributions are well separated from each other. By fitting the
distributions with Gaussian functions, we obtain the mean and width
for each energy which are used to charaterize the detector response.

\begin{figure}[!th]
\centering \includegraphics[width=0.55\textwidth]{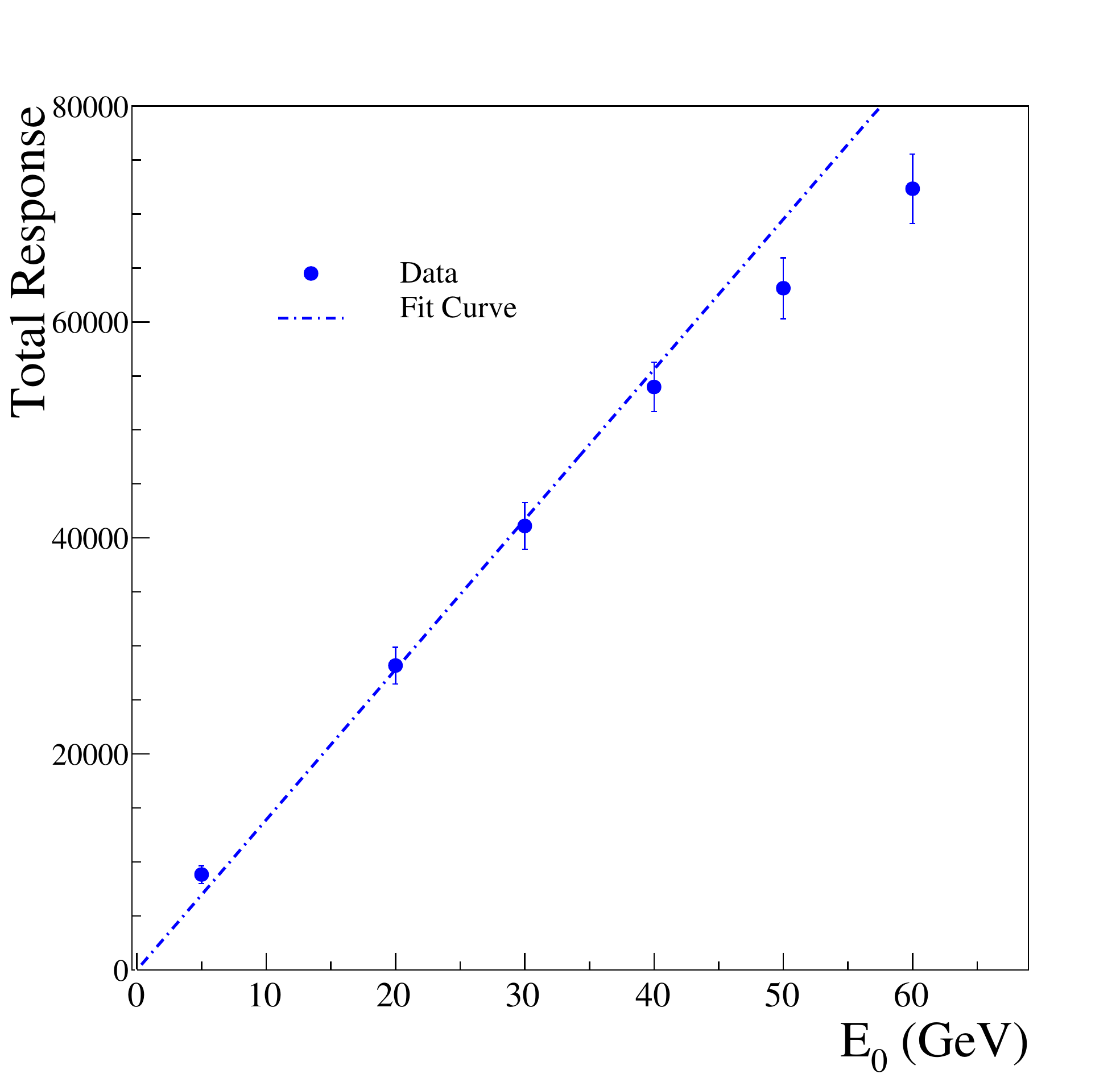}
\caption{Total measured energy deposition for events. The error bars
  represent statistical uncertainties. A linear fit of the form $f = (1.390\pm 0.038) \cdot 10^{3} 
E_{0}{/\rm  GeV}$ can explain the data well.}
\label{calibrationSubsampleMethod}
\end{figure}

Fig.~\ref{calibrationSubsampleMethod} shows the mean of the total
response as a function of the incident energy of electrons. 
A linear response is found, except for the highest two energyes (50 and 60 GeV), where a saturation effect due to the limited limited dynamic range of the
readout electronics has been indentified during the data analysis. A linear fit to the data points below 50 GeV yields: 
${\rm Response} = (1.390\pm 0.038) \cdot 10^{3} 
E_{0}{/\rm  GeV}$, where $E_{0}$ is the incident beam energy.

\subsection{Energy resolution}

\begin{figure}[!th]
\centering \includegraphics[width=0.55\textwidth]{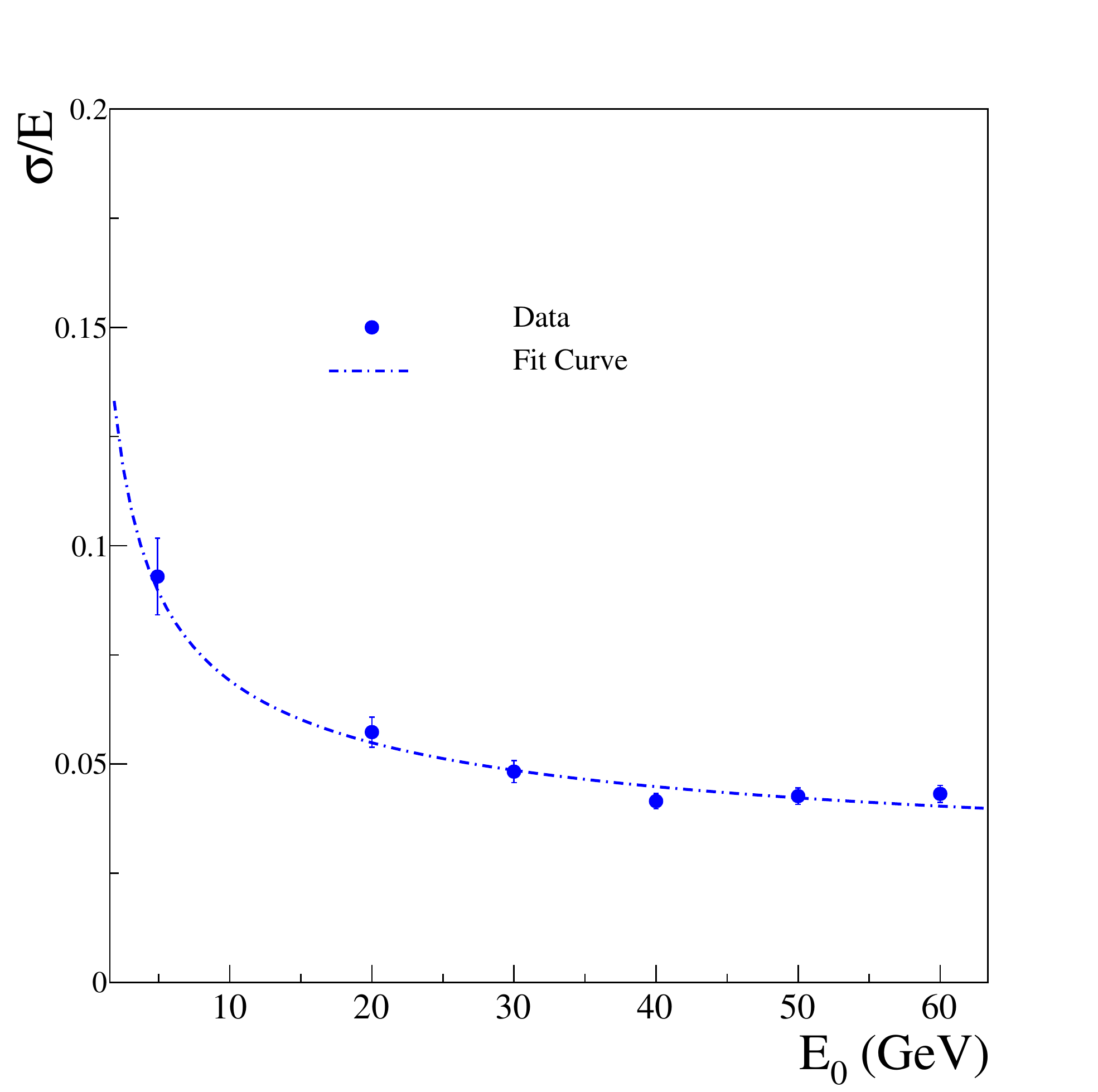}
\caption{Measured energy resolution of the Si-W calorimeter prototype. 
The error bars represent statistical uncertainties. 
A fit for the form,
$\frac{\sigma}{E} = a  \oplus \frac{b} {\sqrt {E_0/\mathrm{GeV}}}$, with 
$a=0.020\pm 0.0038$ and $b=0.1536\pm 0.023$, fits the data points very well.
}
\label{ResolutionSubsampleMethod}
\end{figure}

The energy resolution is one of the most important parameters to
characterise a calorimeter. For a given incident energy, the 
resolution is defined as the ratio of the width~($\sigma$) to the mean~($E$) of the
measured energy or the total energy response.  
Figure~\ref{ResolutionSubsampleMethod} shows the measured
energy resolution~($\sigma/E$) of the calorimeter, plotted as a function of
the incident energy. With the increase of the incident energy,
resolution decreases. The measured energy
resolution includes the combined effects of the structure of the calorimeter,
the statistical fluctuations of the shower development, and contribution from
electronic noise. The energy dependence of the resolution of a calorimeter is normally
expressed in terms of the following empirical formula~\cite{Fabjan, calice-4}: 
\begin{equation}
\label{eq3}
\frac{\sigma}{E} = a  \oplus \frac{b} {\sqrt {E_0/\mathrm{GeV}}}, 
\end{equation}
where $a$ and $b$ are the fit parameters. The coefficient $b$ gives
the contribution from fluctuations in the shower development, which
have a characteristic $\frac{1}{\sqrt{E}}$ dependence, while $a$
quantifies the energy-independent contribution. The constant term
specifies the imperfections in the construction of the calorimeter and
is sensitive to non-uniform response of the detector. It is crucial to
have a low value of the constant term (about 1\%)  to achieve good energy
resolution at high energy. 
A fit of this form to the calorimetric data of
Fig.~\ref{ResolutionSubsampleMethod} yields, 
$a=0.02\pm 0.003$ and $b=0.1536\pm 0.023$. These coefficients, 
expressed in terms of percentage are: $\sigma/E = 15.36/\sqrt{E_0(\mathrm{GeV)}} \oplus 2.0$ (\%).
This effect has been studied using GEANT4 simulation 
with an approximate detector geometry. The measured resolution is 
close to what has been estimated from the simulation.
It is worth mentioning here that the constant term 
$a$ of 2\% can be 
improved by making the calorimeter more compact, by the use of 
improved readout electronics and with proper relative
calibration of the individual pads.

\section{Summary}

A prototype Si-W calorimeter has been designed, constructed and tested
using pion and electron beams at different energies at the
CERN-SPS. The calorimeter comprises of 19 layers of silicon pad and tungsten layers placed
alternating with each other. The silicon pads are of 1~cm$^2$ area and
300~$\mu$m thick, and the tungsten plates are of 1$X_0$ thick. 
For the readout the MANAS ASIC has been used, which has a limited dynamic range. The
calorimeter was tested with 120GeV pion beams to understand the
behaviour of minimum ionising particles. Electron beams of energy 5 to 60 GeV
provided the electromagnetic showers for each silicon
layer of the detector. Longitudinal shower profiles have been obtained
from which the position of the shower maximum has been extracted. The total
response for different incident energy electrons has been
obtained. A linear behaviour of measured total energy with that of
incident energy ensures satisfactory calorimetric performance. The dependence of 
the measured energy resolution on the incident beam energy $E_0$ can be 
characterised as 
$\sigma/E = 15.3/\sqrt{E_0(\mathrm{GeV)}} \oplus 2$ (\%). 
These results are comparable to the ones obtained by the CALICE
collaboration~\cite{calice-1} for an equivalent
silicon-tungsten sampling calorimeter.

\acknowledgments

We acknowledge valuable discussions held with Yogendra P. Viyogi,
Sinjini Chandra, Premomoy Ghosh, and Subhasish Chattopadhyay. We 
thank Bharat Electronics Limited, Bangalore for providing the silicon
pad detectors. We thank the CERN SPS crew for providing excellent
quality beam for the detector tests
and ALICE-FOCAL collaboration for the support during the tests.


\end{document}